# Energy-efficient topology to enhance the wireless sensor network lifetime using connectivity control

**Meysam Yari**[*1]**, Parham Hadikhani**[2]**, Zohreh Asgharzadeh**[3]

## Abstract

Wireless sensor networks have attracted much attention because of many applications in the fields of industry, military, medicine, agriculture, and education. In addition, the vast majority of researches have been done to expand its applications and improve its efficiency. However, there are still many challenges for increasing the efficiency in different parts of this network. One of the most important parts is increasing the lifetime of the wireless sensor network. In this paper, using topology control, the threshold for the remaining energy in nodes, and two of the meta-algorithms include SA and VNS, we increase the energy remaining in the sensors. Moreover, using a low-cost spanning tree, we create an appropriate connectivity control among nodes in the network in order to increase the lifetime of sensor network. The results of simulations show that proposed method improves the lifetime of sensor and reduces the energy consumed.

## 1   INTRODUCTION

Wireless sensor networks are one of the kinds of wireless ad hoc network that today have many applications in the fields of industry, military, medicine, education and etc. These networks consist of tens to thousands of wireless nodes without monitoring [2], which are easily deployed in different environments and defined specific functions for nodes. One of the advantages of the

Meysam Yari
meysam.yari68@yahoo.com

[1]Department of Computer Engineering, Islamic Azad
University Shabestar Branch, Shabestar, East Azerbaijan, Iran
[2] Department of Computer Engineering, Pasargad Higher Education Institute, Shiraz, Fars, Iran
[3] Department of Computer Engineering, Payame Noor University Germi Branch,
 Germi, Ardabil, Iran

wireless sensor network is the high development speed due to its simplicity and low cost of implementation and automatic configuration. However, because of having features such as the poor battery in sensor nodes, it faces challenges [3]. Many studies have been carried out by researchers on the characteristics of this network to improve many of its capabilities. Because of the importance of this network in most areas, we decided to find a solution to increase the lifetime of sensor nodes by reducing energy loss. Many )heuristic and meta-heuristic( methods have been proposed to increase the lifetime of wireless sensor networks, such as clustering of sensor nodes, connectivity control, mobile-Sink and etc. Energy consumption in wireless sensor nodes has a direct relationship with the lifetime of the wireless sensor network so that if we reduce the process of energy consumption in the sensor nodes, we can claim that we are able to increase the network lifetime. One of the methods increases network lifetime is connectivity control in the wireless sensor network [1]. In this paper, we are looking for creating a Spanning tree that can increase the lifetime of the wireless sensor network. A way to achieve this goal is to use mobile-Sink and a Spanning tree from sensor nodes in the network. Sensor nodes send their received data to the base station by one or several sinks for analysis and statistical reports. Simulation results show that the lifetime of the wireless sensor network has improved in this study compared to [1]. In the first part of the article, we discuss the introduction of wireless sensor networks. In the second part, we introduce the related works. In the third part, the proposed method is presented and in the fourth section, simulation of the proposed method is expressed and finally, in the final section, the conclusion and future works are expressed.

## 2  RELATED WORK

The wireless sensor network is categorized in the wireless ad hoc network, which has attracted a lot of attention in terms of academic and industrial in recent years [1]. Despite the many advances made in the wireless sensor network, we still face challenges including energy consumption, routing, scalability, security, and fault tolerance. Routing and finding the most appropriate path among nodes of wireless sensor network have many benefits including connectivity control, increasing throughput, and eventually increasing the lifetime of the wireless sensor network. In [4], the "maximum lifetime algorithm" has been proposed where nodes with a greedy policy are added to the routing tree one by one. Furthermore, in this paper, the proposed algorithm seeks to increase the lifetime of the wireless sensor network without knowing about the queries and their production rates. Only nodes can be added to the routing tree that they are able to increase the lifetime of the network. In [5], the LEACH algorithm has been expressed. The algorithm is a hierarchical algorithm that at one level, it has a number of nodes as head clusters and at the next level, the nodes that do not belong to any of the head clusters, are members of these clusters. One of the advantages of this method is load balancing increase among all network nodes. In [6], two PEDAP and PEDAP - PA protocols are proposed and the goal is to increase the lifetime of the wireless sensor network and the balanced energy consumption in each node. The results show that the proposed protocols have better performance than LEACH and PEGASIS protocols to increase the lifetime of the wireless sensor network. In [7], linear programming 0-1 is used to find the optimal mapping among the members. After that, a two-dimensional genetic algorithm is used for optimal routing among members. Finally, a two-phase focused communication protocol is used to support the "maximum value of shortest path" algorithm. In [6], the problem of balancing energy is formulated as the problem of optimizing the allocation of data transmission by combining the idea of network division based on CORONA and the hybrid routing strategy. The proposed EBDG protocol has better output compared to Multicast transmission schemes, direct transfer, and cluster rotation. In [8, 9], they are looking for the load balanced consumption to aggregate of the sensor network data. The solution is followed by a comparison among data transfer as hop by hop and direct transfer from the nodes. Thus, two RLN and GCN models are introduced to reduce energy consumption in the network. In [10], the proposed algorithm combines the collected data with the desired routing and presents a smooth approximation function to optimize the problem. As a result, the network lifetime increases by routing and maximizing data aggregation. It also affects network traffic reduction. In [11], a new model is proposed to predict the lifetime of a wireless sensor node on the basis of the Markov model (MPLM). Additionally, TCAMPLM is provided by adjusting the transmit power of sensor nodes to keep energy in the nodes. In [12], the mobile service computing algorithm is proposed to solve the problem of connectivity control in the wireless sensor network. In [13], a connectivity control algorithm based on the learning automata called LBLATC is proposed. The learning automata chooses the appropriate transmission range of nodes to use the reinforcement signal generated by sensor nodes. The simulation results show that the expressed protocol has proper performance. In [14], an algorithm is presented to improve the lifetime in wireless sensor networks. This algorithm initially detects the dynamic hole then bypass the hole so that the nodes around the hole consume less energy. In [1], by using a greedy policy and dynamic programming, an innovative connectivity control algorithm (MLS) is proposed that use the mobile sink. The purpose of this method is to increase the minimum node energy in the wireless sensor network which leads to maximizing the lifetime of the wireless sensor network. This method has better performance in increasing the lifetime of the wireless sensor network compared to previous methods.

## 3  Proposed Method

In this paper, we are looking for enhancing the lifetime of the sensor network by using connectivity control of sensor nodes and considering three limits based on the remaining energy of the sensor nodes. In fact, in this network, a certain threshold for this local search is obtained and the changes in communications among nodes are happening when their energy level reaches this threshold. In the following, we use the meta-heuristic algorithms which include variable neighborhood search (VNS) [15] and simulated annealing (SA) [11] to achieve a spanning tree that maximizes the minimum lifetime of sensor networks. We assume that a number of wireless sensor nodes are located in an environment with a mobile sink. The mobile sink gathers the data obtained by sensor nodes along its path. Sensor nodes in this research are divided into two categories: anchor sensor nodes and normal sensor nodes. Anchor sensor nodes are nodes close to the path of the mobile sink, and the normal nodes are nodes that farther away from the path of the mobile sink. The normal nodes send received data from the surrounding environment or their next normal node as single or multiple order to anchor nodes [1]. The anchor nodes send received data from the surrounding environment, along with the data which is received from their previous normal nodes to the mobile sink when it close to anchor nodes. This connection creates connectivity between sensor nodes and the mobile sink. We want to create a spanning tree from the wireless sensor nodes with the root of the mobile sink that contains all network nodes. Our objective is to maximize the minimum lifetime of the sensor network so that the death of the first sensor node occurs later.

### 3.1 Energy consumption model

Sensor nodes regardless of what roles they can play in a wireless sensor network, if they have enough energy, the sensor nodes can receive data from their surrounding environment and send the received data to the next node that is closer to the path of the mobile sink according to the algorithm available in the network. They can also receive data from their neighbors depending on the defined methods in that network, and send the received data along with their data to the next sensor node. Generally, the sensor nodes consume more energy at the time of sending and transmitting the data to the next node. In [١٦], according to the first - order radio model, the cover radius is defined based on the energy required in the free/ multipath space as follows:

$$Em = \{(E(elect) + E(da) \times l) + (Eamp \times l \times d^{\alpha})\} \quad (1)$$

In equation (1), *Em* is the energy consumed at the sensor node and *E*(elect) is the energy emission for transmission and receives data. *E*(da) is the energy consumed to aggregate sensor node data and also *Eamp* is the emission of energy for the amplifier (depending on the free space $\alpha - 2$ and Multi-path fading $\alpha - 4$). *I* is the length of the message and d is the distance of the current node from its neighbor node. Thus energy consumption in this research is based on equation (1). The most important part of our research is the remaining energy of the nodes. Each sensor node can be a bottleneck to increase the lifetime of the wireless sensor network. As a consequence, we chose the connectivity control based on the minimum spanning tree to increase the lifetime of the wireless sensor network.

### 3.2 Proposed algorithm

The sensor nodes are randomly distributed in a 1000 meters at 1000 meters environment and the sink node moves in a predefined path. Each node locally receives its distance from neighbors by using the Hello message. Nodes that are close to the path of the mobile sink known as anchor nodes and nodes that farther away from the path of the mobile sink are called non-anchor nodes (normal nodes). Anchor nodes are responsible for sending their data and data of non-anchor nodes to the mobile sink. The relative load of sensor nodes is based on:

$$\text{Load}\_m(v) = \frac{(C_{rx} + C_{tx}(r(v))) \times q(v) - C_{rx}}{e(v)} \qquad (2)$$

That is the cost of energy for the remaining energy in the sensor node, and we are looking for minimizing the maximum relative load of the sensor nodes:

$$\text{minimize} : \max\{l(v) \mid v \in V\} \qquad (3)$$

Thus, nodes that reduce the load of our spanning tree are added to the tree (Figure 1). Initially, we start from the root node (the sink). After the anchor nodes are identified and connected to the root node, in the next step, the relative load of neighbor nodes is calculated according to formula 2 if they were not previously members of the spanning tree and their remaining energy has not reached by the threshold value. A node that has the lowest relative load (initial selection phase) [1] is selected as the candidate nodes. The amount of relative load is obtained on the basis that if the candidate node is to be attached to the tree, its cost will be relative load. In the next step (final selection phase), according to nodes that have been selected in the previous step, the load of candidate nodes in the spanning tree are sorted in ascending order and the largest amount of load is compared with the largest amount of load of the other candidate nodes. Then, the lowest amount is added as the selected node to the spanning tree (Figure 2). Three limits are defined for network nodes which include threshold, warning, and death. In each round of energy simulation, which each node needs to send its data to the next node, it is specified in a table. If the remaining energy in a node reaches the threshold (twice the minimum energy stored for each node), we enter the local search step, where that node is labeled and the children of that node should be separated from their parent and the parent selection process should be re-established for them. It leads to reduce the speed of energy loss of the labeled node, and this helps to increase the lifetime of the network throughout the network. The reason for the double value is that if the remaining energy of the node reaches twice its minimum stored energy, it will have the opportunity to increase its lifetime by entering a local replacement phase. This means that with two sink moves, the node will not be able to send/receive data in the network and will be removed from the data transfer in the network. But if we can make this node consume less energy in each sink move, we help to increase its lifetime. The warning threshold means that if the remaining energy in the node is three times the minimum energy stored for the node, we will enter the phase of finding new neighbors. If the remaining energy in the node is less than the minimum amount of energy stored for the node, the threshold condition of the death of node is fulfilled and that node will be removed from the data transfer in the network.

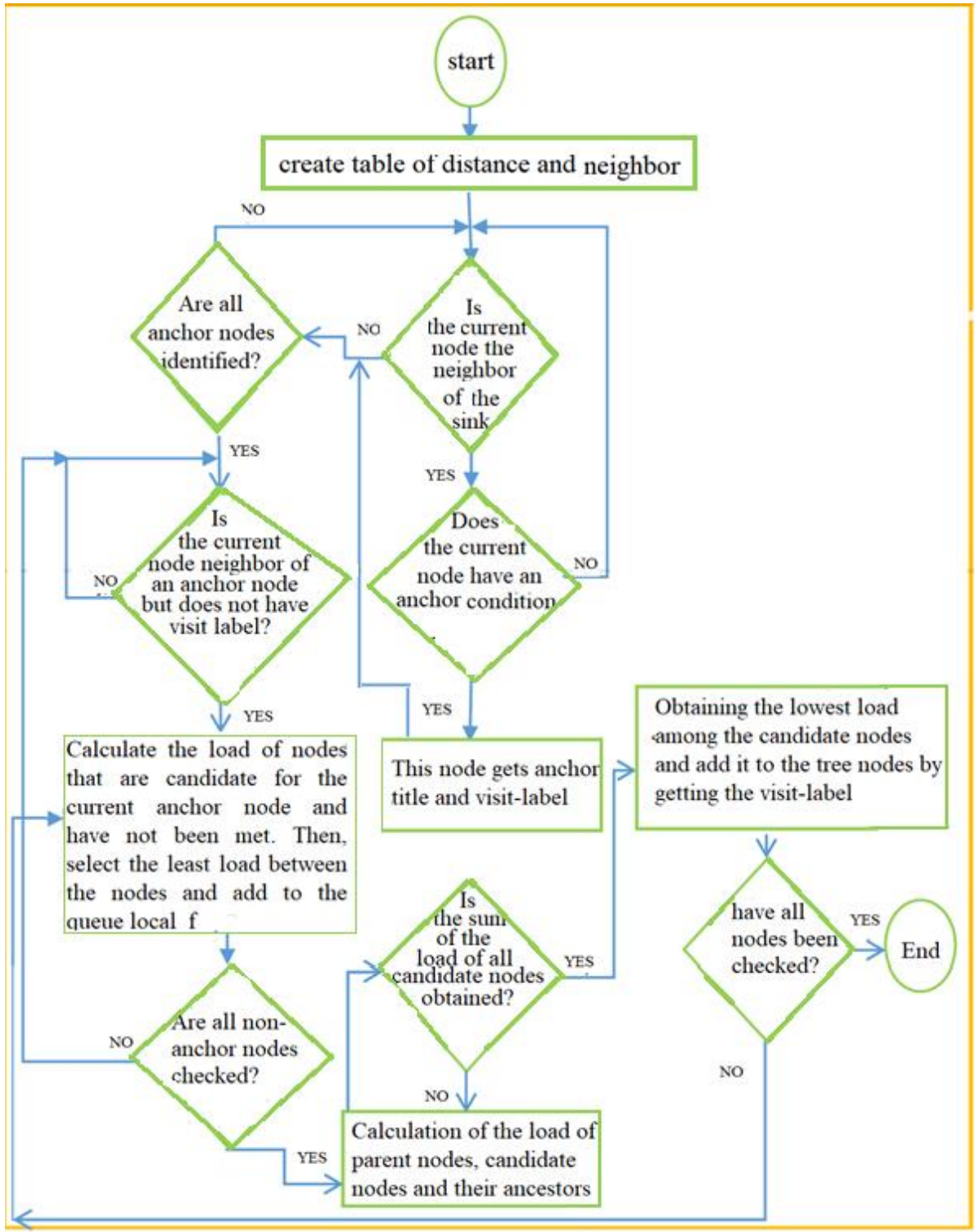

Figure 1. Steps to create spanning tree

The reduction of the number of child nodes of a node means shifting the load to subtrees. In the local search phase, the child nodes are separated from that node and a new parent is chosen for them. Thus, the energy consumption of the desired node is reduced. Meanwhile, if the node x that has reached the threshold has only one child and no child in its neighborhood, in the local search time, the child node is separated from node x and is detached from the network. To prevent being isolated from a node on the network, we use the warning threshold. If a node is placed in such a case, another node is introduced as the parent node of that node.

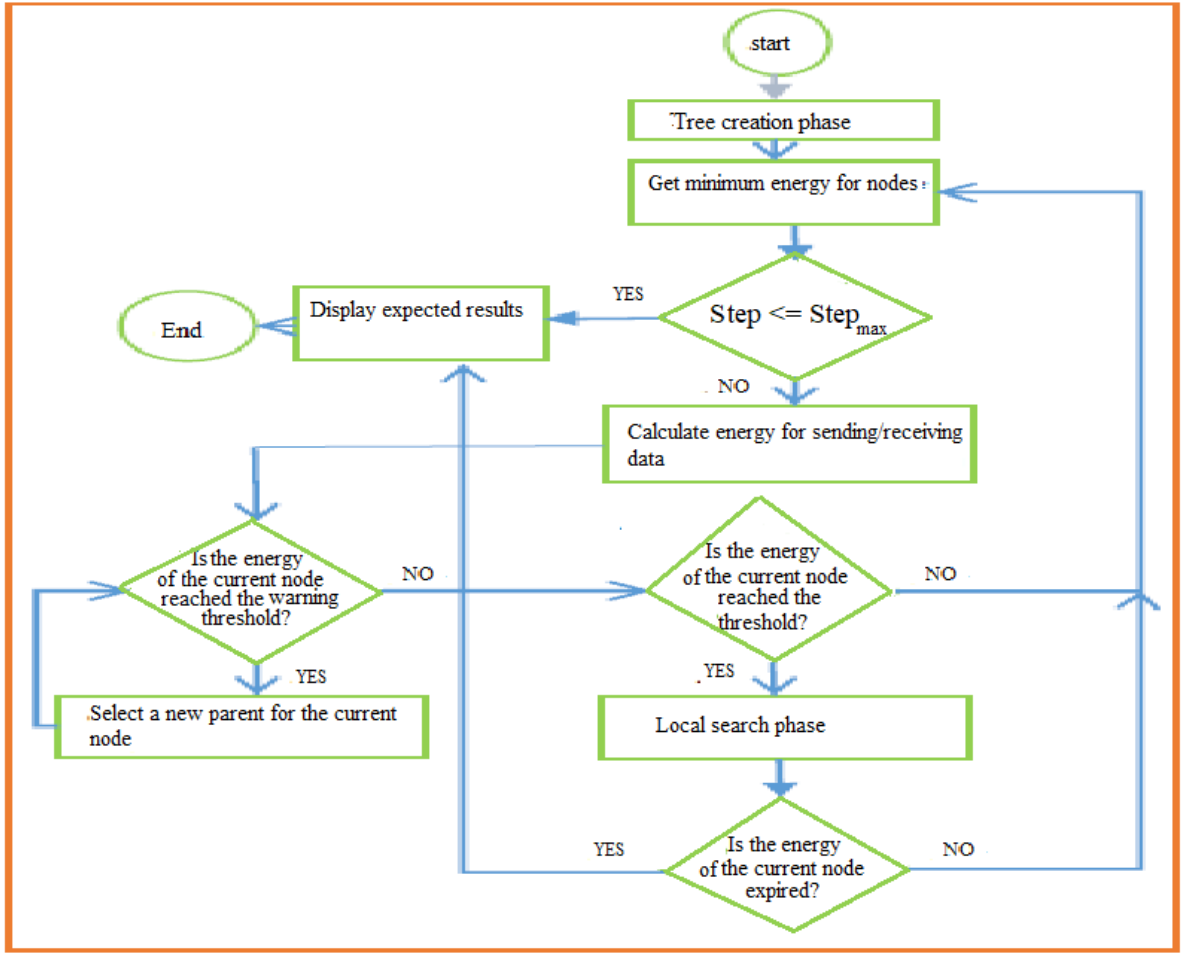

Figure 2. Proposed Method

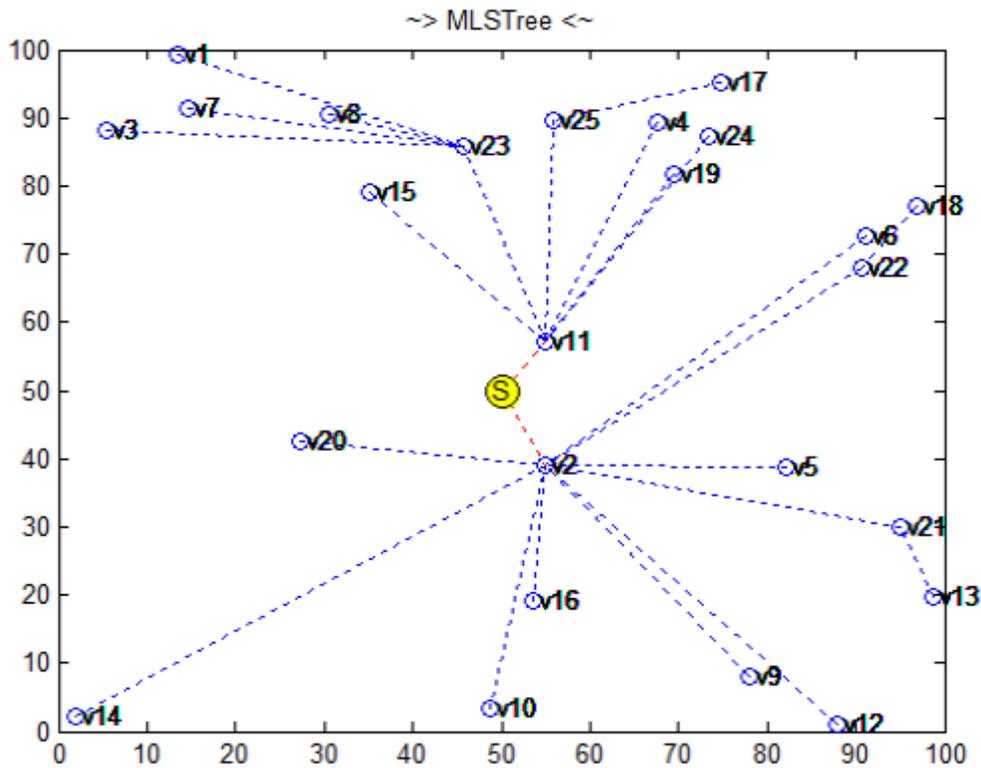

Figure 3. Completed network with the proposed method

As stated earlier, the warning threshold and the threshold are used to if after making routing spanning, the traffic of sending packets on a node increases, they can divide load of that node over the rest of the nodes in trees. By doing this work, it's possible to claim that the lifetime of the network has increased over before. To obtain the remaining energy, we used simulated annealing [16] and variable neighborhood search [3]. We are looking for a tree that has less construction cost. The purpose of the cost is to select the node with the minimum load from the tree nodes with the highest load. As stated earlier, the warning threshold and the threshold are used to if after making routing spanning, the traffic of sending packets on a node increases, they can divide the load of the node over the rest of the nodes in trees. By doing this work, it's possible to claim that the lifetime of the network has increased over before.

```
initialization
Number of Population;
Number of Move toward Neighbors;
initial temperature and find temperature and T = T0 and α is reduction rate;
initialize best_sol.cost = Inf;
Pop = Generate Population(with random solution);
if Pop.cost <= best_sol.cost then
    best_sol = Pop;
repeat until stopping condition is met(T <= T_F)
    create and evaluate new_solution for Number of popluation and moving;
    new_pop = create Neighbor for every member of population;
    for i ← 1 to number of pop do
        if new_pop.cost <= Pop.cost
            Pop = new_pop;
        elseif rand < TempFunc(new_pop.cost, Pop.cost, T)
            bes_sol = Pop;
        store best_sol;
T = α × T;
```

Algorithm 1. The pseudo code of the proposed method with SA

First, we need an initial solution that we consider a neighborhood relationship among the nodes. We consider the cost of building a tree as the initial solution and enter the simulation annealing phase. For all the processors that we consider as the set of responses, we make changes using the neighborhood function in the generated answers. Afterward, we compare the cost of the obtained response with the cost of the best answer so far. If the answer (minimum) is better than the best answer, the answer is replaced by the best answer. Otherwise, by creating a random number and a Boltzmann probability function, we are looking to accept the bad answer as well. If this is not true, the best answer will not change (see the Algorithm 1).

```
Initialization
    select the set of neighborhood structures
    N_k
    find an initial solution x
repeat until stopping condition is met
    set K = 1;
    repeat until K = K_max
        1- Do_Shaking: Generate a random point x` in N_k(x);
        2- Local_search: X`` is optimum obtained;
        3- Move / not Move:
            - if x`` is better than x, then x = x`` and K = 1;
            - else K = K + 1;
```

Algorithm 2. Pseudo code of VNS [3]

The next algorithm is Variable Neighborhood Search [3] (Algorithm 2). The algorithm uses two parameters: "vibration " and " local search ". Vibration generates diversity and local search look for the most appropriate answer. The vibration section creates a fundamental change in the initial response. Since it takes a lot of time to search for the entire problem space, using vibration, we make diversity in the answer to almost make sure that we check every state of the answer.

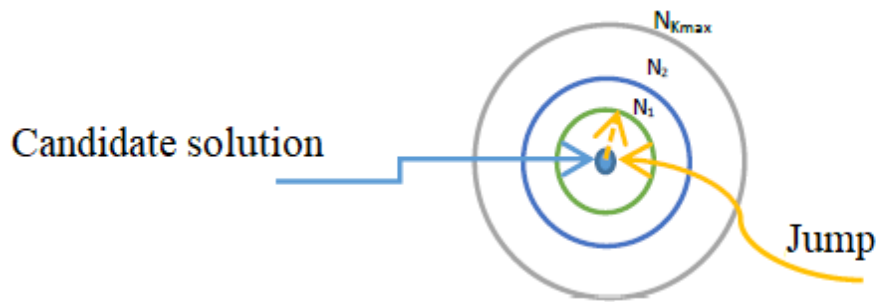

Figure 4. Variable Neighborhood Search [3]

In the local search phase, if an answer is found better than the current solution, the current solution will replace it. Otherwise, it goes to other areas to compare their answers to the current solution. We do this work by the end of the specified time in local search. Finally, the output of this stage is compared with the best answer that already existed. If the solution is better than the current solution, then the solution is replaced by the current solution. Otherwise, the vibration is performed on the last answer obtained from the previous step to check again the answers of that range.

## 4  Simulation and results

In this section, in order to simulate the proposed method and compare it with [1], we use the ns2 simulation tool with mannasim. Ns2 is an event-based software that is presented as open source. In this simulation, it is assumed that the simulation environment is 1000 meters at 1000 meters and consists of 100, 130, 160, 190, and 190 wireless sensor nodes along with a mobile sink. For routing, the AODV protocol has been used and the radio antenna is omnidirectional. All parameters are presented in Table 1.

Table 1: Parameters

| Parameter | Value |
|---|---|
| The primary energy of each node | 0.2 J |
| Area of simulation | 1000m*1000m |
| Number of sink | 1 |
| MAC type | Mac/802-11 |
| Clustering algorithm | LEACH |

To achieve the simulation results, the number of rounds has been executed 25 times that each execution has 1200 runs. Next, we obtained the mean of these 25 executions and the mean values are recorded in the results.

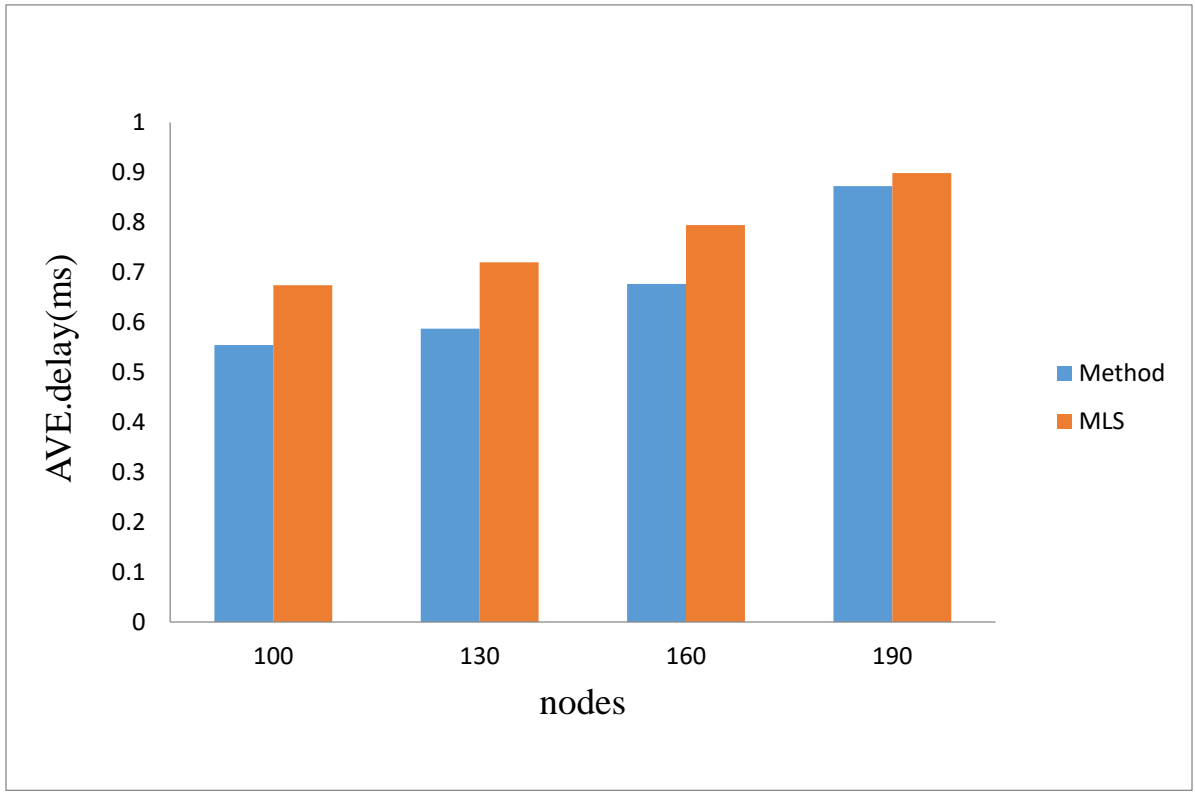

Figure 5. The comparison of the average end-to-end delay between the proposed method and [1] in the sensor network

Fig. 5 shows the mean end-to-end delay among nodes in milliseconds. After 22 execution for 100 nodes in the proposed method, the average delay is 0.5543 while in method [1] is 0.67424. When the number of sensor nodes is 130, the average delay in the proposed method is 0.58725, however in method [1] is 0.719874. In addition, when the number of sensor nodes is 160, the average delay is 0.676539 compared to 0.794338 in method [1]. Eventually, when the number of sensor nodes is 190, the average delay in the proposed method is 0.8725 as opposed to 0.898538 in [1].

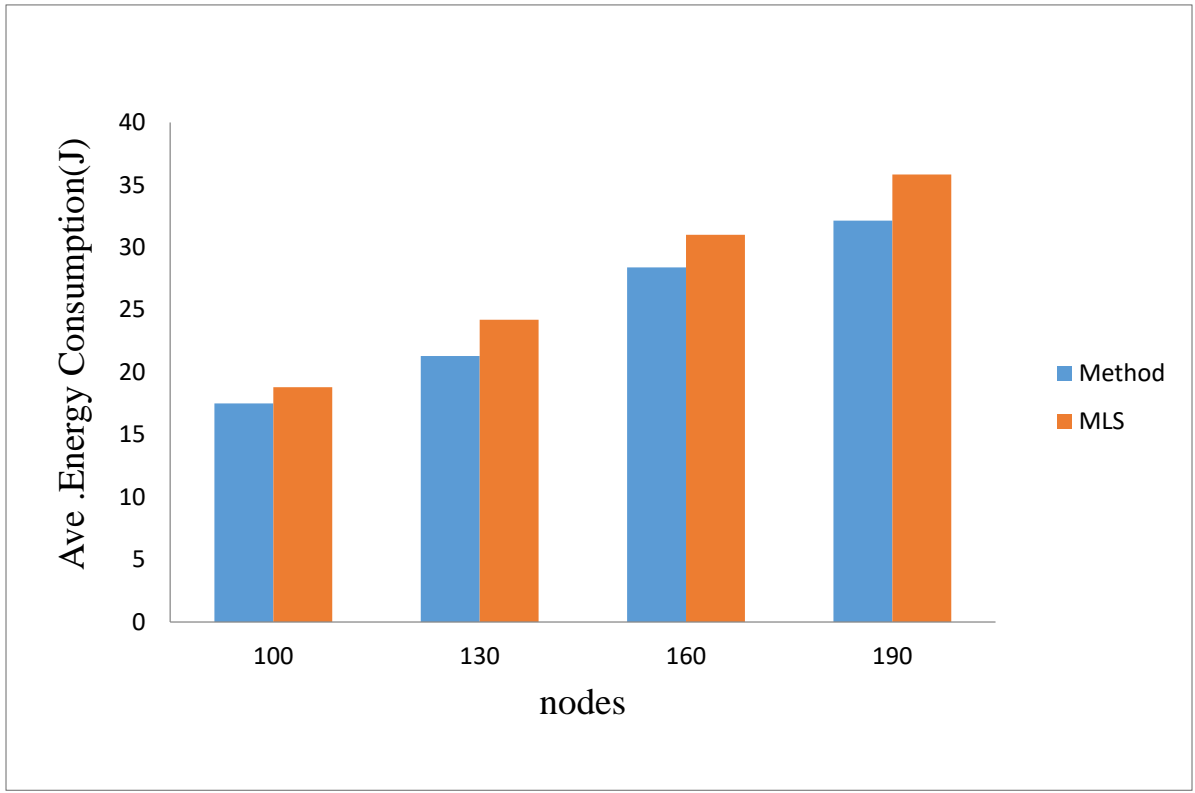

Figure 6. The comparison of the average energy consumed between the proposed method and [1] in the sensor network

In Figure 6, we examined the energy consumed in the sensor nodes. For 100 nodes, the average energy consumption after 25 runs in the proposed method is 17.5 J, whereas in [1] is 18.8 J. For 130 sensor nodes, the average energy consumption in the proposed method is 21.3 J, compared to [1] which is 24.2 J. For 160 and 190 sensor nodes, the proposed method consume less energy than method in [1], 28.39 J and 32.14 J respectively compared to 31 J and 35.83 J.

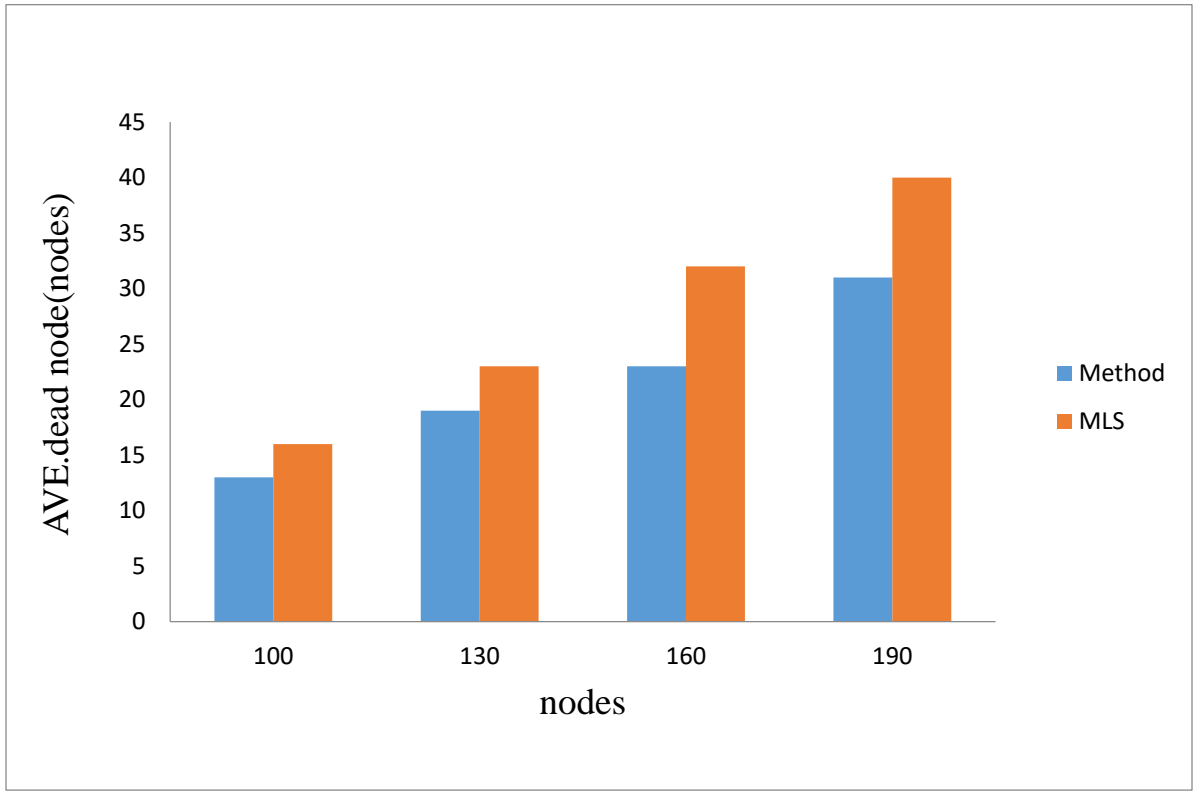

Figure 7. The comparison of the average lifetime of the sensor nodes between the proposed method and [1] in the sensor network

Figure 7 reveals the lifetime of wireless sensor nodes based on the dead node between our method and the proposed method in [1]. When the number of nodes is 100, the number of dead nodes in our method and [1] are 13 and 16 respectively. Moreover, when the number of nodes is 130, the number of dead nodes in our method is 19, slightly less than 23 dead nodes in [1]. In the same way, For 160 and 190 sensor nodes, the number of dead nodes in our method are 23 and 31 in turn, as opposed to 32 and 40 dead nodes in [1] respectively.

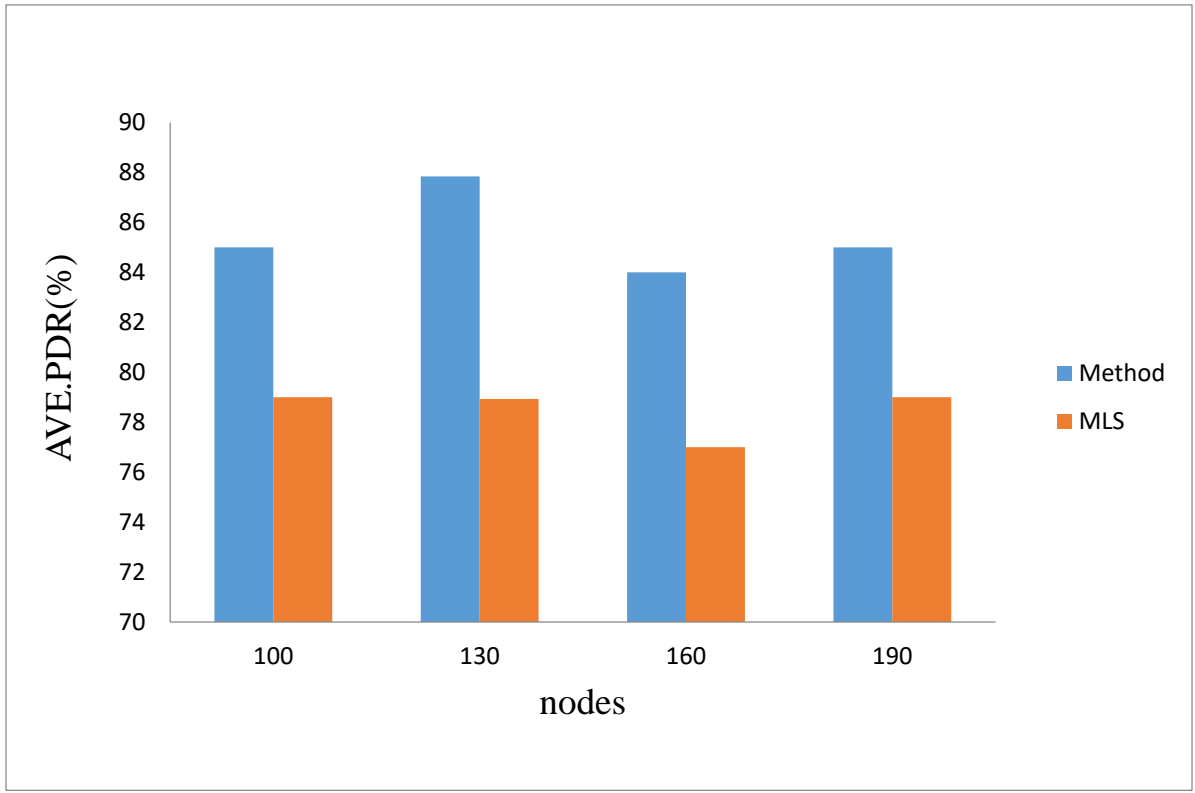

Figure 8. Comparison of average packet delivery rates among network nodes between the proposed method and [1]

Figure 8 indicates the comparison of average packet delivery rates among network nodes for the proposed method and [1]. When the number of nodes is 100, Average packet delivery rate in the proposed method is 85% compared to [1] which is 79%. When the number of nodes is 130, the average packet delivery rate in the proposed method is 81.84% as opposed to 78.93% in [1]. For 160 and 190 sensor nodes, the average packet delivery rates in the proposed method are 84% and 85% in turn, while in [1] are 77% and 79% respectively.

## 5   Conclusions and future works

Considering the many applications of wireless sensor networks and the great passion for using it, it still faces challenges including energy consumption. In this study, we developed load balancing by using proper connectivity control between the sensor nodes and the definition of three limits including thresholds, warnings, and deaths for sensor nodes. Moreover, with the help of two SA and VNS meta-algorithms, we found the lowest cost spanning trees for sensor nodes. The results of the simulation show that the lifetime of the sensor network in the proposed method has improved and enhanced compared to [1]. As the future works, in order to evaluate the lifetime of sensor nodes before forming the connectivity formation, we can use the MPLM method in [31]. In the proposed method, we can have the durability or persistence of sensor nodes actively in the network.

As a result, the connectivity is established among sensor nodes which includes the best mode and network status and improves the network lifetime. The second suggestion is using an information packet that is exchanged among all sensor nodes and includes their current status (active/deactivated). This packet has the status of all sensor nodes and this information can be used to better routing of the mobile sink.

# 6 REFERENCES

1. Zhao, H., et al., *Energy-efficient topology control algorithm for maximizing network lifetime in wireless sensor networks with mobile sink.* Applied Soft Computing, 2015. **34**: p. 539-550.
2. Vecchio, M. and R. López-Valcarce, *Improving area coverage of wireless sensor networks via controllable mobile nodes: A greedy approach.* Journal of network and computer applications, 2015. **48**: p. 1-13.
3. Yetgin, H., et al., *A survey of network lifetime maximization techniques in wireless sensor networks.* IEEE Communications Surveys & Tutorials, 2017. **19**(2): p. 828-854.
4. Liang, W. and Y. Liu, *Online data gathering for maximizing network lifetime in sensor networks.* IEEE transactions on mobile computing, 2006. **6**(1): p. 2-11.
5. Heinzelman, W.B., A.P. Chandrakasan, and H. Balakrishnan, *An application-specific protocol architecture for wireless microsensor networks.* IEEE Transactions on wireless communications, 2002. **1**(4): p. 660-670.
6. Tan, H.Ö. and I. Körpeoğlu, *Power efficient data gathering and aggregation in wireless sensor networks.* ACM Sigmod Record, 2003. **32**(4): p. 66-71.
7. Gao, S., H. Zhang, and S.K. Das, *Efficient data collection in wireless sensor networks with path-constrained mobile sinks.* IEEE Transactions on Mobile Computing, 2010. **10**(4): p. 592-608.
8. Zhang, H. and H. Shen, *Balancing energy consumption to maximize network lifetime in data-gathering sensor networks.* IEEE Transactions on Parallel and Distributed Systems, 2008. **20**(10): p. 1526-1539.
9. Zhang, H., H. Shen, and Y. Tan. *Optimal energy balanced data gathering in wireless sensor networks.* in *2007 IEEE International Parallel and Distributed Processing Symposium.* 2007. IEEE.
10. Hua, C. and T.-S.P. Yum, *Optimal routing and data aggregation for maximizing lifetime of wireless sensor networks.* IEEE/ACM Transactions on Networking (TON), 2008. **16**(4): p. 892-903.
11. Hao, X., et al., *Topology control game algorithm based on Markov lifetime prediction model for wireless sensor network.* Ad Hoc Networks, 2018. **78**: p. 13-23.
12. Hou, J. and Y. Zhang, *Mobile-Service Based Approach for Topology Control of Wireless Sensor Networks.* Wireless Personal Communications, 2018. **102**(2): p. 1839-1851.
13. Javadi, M., et al., *Learning automaton based topology control protocol for extending wireless sensor networks lifetime.* Journal of Network and Computer Applications, 2018. **122**: p. 128-136.
14. Hadikhani, P., et al., *An energy-aware and load balanced distributed geographic routing algorithm for wireless sensor networks with dynamic hole.* Wireless Networks: p. 1-13.
15. Hansen, P., et al., *Variable neighborhood search: basics and variants.* EURO Journal on Computational Optimization, 2017. **5**(3): p. 423-454.
16. Kirkpatrick, S., C.D. Gelatt, and M.P. Vecchi, *Optimization by simulated annealing.* science, 1983. **220**(4598): p. 671-680.